\documentclass[twocolumn,trackchanges]{aastex61}
\usepackage[utf8]{inputenc}
\usepackage{units}
\usepackage{color}
\usepackage{calc}
\usepackage{graphicx}
\usepackage{enumerate}
\usepackage{mathrsfs}
\usepackage{amsmath}
\usepackage{amssymb}
\usepackage{amsfonts}
\usepackage{bm}

\begin{document}
\def\calE{{\cal E}}
\def\calL{{\cal L}}
\def\bk{{\bf k}}
\def\bq{{\bf q}}
\def\bv{{\bf v}}
\def\bu{{\bf u}}


\shorttitle{Generation of suprathermal electrons}
\shortauthors{Tigik, Ziebell and Yoon}

\title{Generation of suprathermal electrons by
collective processes in collisional plasma}
\author{S. F. Tigik}
\email{sabrina.tigik@ufrgs.br}
\affiliation{Instituto de F\'{\i}sica, Universidade Federal
  do Rio Grande do Sul, 91501-970 Porto Alegre, RS, Brazil}
\author{L. F. Ziebell}
\email{luiz.ziebell@ufrgs.br}
\affiliation{Instituto de F\'{\i}sica, Universidade Federal
  do Rio Grande do Sul, 91501-970 Porto Alegre, RS, Brazil}
\author{P. H. Yoon}
\email{yoonp@umd.edu}
\affiliation{Korea Astronomy and Space Science Institute,
  Daejeon,Korea}
\affiliation{Institute for Physical Science \& Technology,
University of Maryland, College Park, MD 20742, USA}
\affiliation{School of Space Research, Kyung Hee University,
  Yongin, Korea}

\begin{abstract}
   The ubiquity of high-energy tails in the charged particle velocity
   distribution functions observed in space plasmas suggests
   the existence of an underlying process responsible for taking
   a fraction of the charged particle population out of thermal
   equilibrium and redistributing it to suprathermal velocity
   and energy ranges. The present Letter focuses on a new and
   fundamental physical explanation for the origin of suprathermal
   electron distribution function in a highly collisional
   plasma. This process involves a newly discovered electrostatic
   bremsstrahlung emission that is effective in a plasma in which
   binary collisions are present. The steady-state electron velocity
   distribution function dictated by such a process corresponds to
   a Maxwellian core plus a quasi-inverse power-law tail, which is
   a feature commonly observed in many space plasma environment. In
   order to demonstrate this, the system of self-consistent particle-
   and wave- kinetic equations are numerically solved with an initially
   Maxwellian electron velocity distribution and Langmuir wave
   spectral intensity, which is a state that does not reflect
   the presence of electrostatic bremsstrahlung process, and hence not
   in force balance. The electrostatic bremsstrahlung term
   subsequently drives the system to a new force-balanced steady
   state. After a long integration period it is demonstrated
   the initial Langmuir fluctuation spectrum is modified,
   which in turn distorts the initial Maxwellian electron
   distribution into a velocity distribution that resembles
   the said core-suprathermal velocity distribution.
   Such a mechanism may thus be operative at the coronal source region,
   which is characterized by high collisionality.
\end{abstract}

\section{Introduction}
\label{intro}

Inverse power-law velocity or energy distributions of charged
particles are either directly observed or inferred in various
regions of the universe accessible to either direct or remote
observations, which includes $\unit[4--5]{MeV}$ protons accelerated
at the heliospheric termination shock and detected by the
{\it Voyager 1} and {\it 2} spacecraft \citep{Stone2008}, tens of
MeV electrons energized at the magnetic-field loop-top X-ray
sources during solar flares \citep{Krucker2014, Oka2015}, energetic
ions and electrons measured in the geomagnetic tail region
during disturbed conditions \citep{Christon1991}, etc.
The solar wind is also replete with background populations of
protons and electrons featuring inverse power-law tail
distributions even in extremely quiet conditions
\citep{Gloeckler2003, Fisk2012, Vasyliunas68, FABMG75, Lin1998}.

In particular, the solar wind electron velocity distribution
function (EVDF) is composed of a Maxwellian core ($\gtrsim 95\%$
of the total density), with energies around $\unit[10]{eV}$, a
tenuous ($4\sim 5\%$) high-energy halo with energies up to
$\unit[10^2\sim 10^3]{eV}$, and a highly energetic ``superhalo''
population with the density ratio of $10^{-9}\sim 10^{-6}$ and
with energies reaching up to $\sim \unit[100]{keV}$~\citep{Lin1998}.
For fast wind, sometimes a narrow beam-like structure, called
the \textit{strahl}, which is aligned with the magnetic field
and streaming in the anti-sunward direction, is also measured
\citep{FABG76, FABGL78, PML99}.

Given the prevalence of non-thermal distributions in nature,
the study of the charged particle acceleration mechanisms
that produce such distributions is of obvious importance
and has a wide-ranging applicability across different
sub-disciplines in astrophysical and space plasma physics.
One of the first kinetic models on how suprathermal electron
populations are generated involves the assumption that a
sub-population of suprathermal electrons in low coronal
regions exists, which is ``selected'' by Coulomb collisions
and interacts with the thermal core and the surrounding
environment in order to form the power-law EVDF at
$\unit[1]{AU}$ \citep{SO79a, SO79b}. Later improved models
generally rely on Coulomb collisional dynamics at the coronal
base and phase-space mapping along inhomogeneous solar
magnetic field lines \citep{L-SHVL97, PML99, PML2001}.
Collisional effects, however, become rather insignificant for
solar altitudes higher than, say, 10 solar radii. In order to
explain the observed quasi-isotropic nature of EVDF near
1 au, wave-particle resonant interaction must be important.
Thus, the collective effects on the EVDF have been considered
with or without other global features \citep{VSLM05, Vocks2012,
PVYZG13, SNYS15, KYCM16b}.

An outstanding issue is whether the suprathermal EVDFs are
generated at the coronal source region in the first place.
This issue may have important ramifications on the
coronal heat flux and inverted temperature profile. If an
enhanced number of high-energy particles is assumed to be
present in the low transition region of the sun, more particles
are capable of escaping the gravitational potential, unleashing
the so-called ``velocity filtration effect,'' which is shown
to produce the observed temperature inversion in the solar
corona, a feature that may be relevant to the coronal heating
\citep{Scudder92a, Scudder92b, Teles15}.
In this regard, \cite{CG14} proposed a scenario in which
electron streams accelerated by nanoflares can lead to the
two-stream instability, and ultimately produce a core-halo
distribution in the inner corona. According to their model,
the core-halo population is simply convected outward along
open field lines while preserving the phase-space properties.

In this Letter we propose an alternative mechanism. This is
not an acceleration in the traditional sense, but rather it
is a mechanism that relies on a new fundamental plasma process
involving the wave-particle interaction in a collisional plasma.
Our theory is based on a recent paper by \cite{YZKS16}, where the
kinetic theory of collective processes in collisional plasmas
was formulated. The problem of combined collisional dissipation
and collective processes had not been rigorously investigated
from first principles in the literature. This is not to say
that collisional dissipation processes or collective processes
are not understood separately. On the contrary, each process is
well understood. Indeed, if one is interested in the situation
where the binary collisional relaxation is dominant, then
transport processes can be legitimately discussed solely
on the basis of the well-known collisional kinetic equation
\citep{Helander2002, Zank2014}. Conversely, if one's concern
is only on relaxation processes that involve collective
oscillations, waves, and instabilities, then there exists a
vast amount of literature on linear and nonlinear theories
of plasma waves, instabilities, and turbulence. It is the
dichotomy that separates the purely collisional versus purely
collective descriptions that had not been rigorously bridged
until \citep{YZKS16}.

Among the findings of \cite{YZKS16} is a hitherto-unknown
effect that came out without any ad hoc assumption. The first
principle equation of this new effect depicts the emission of
electrostatic fluctuations, in the eigenmode frequency range,
caused by particle scattering. This electrostatic form of
``braking radiation'' was appropriately named electrostatic
bremsstrahlung (EB) by the authors of \cite{YZKS16}, which
is not to be confused with a process sometimes known in the
literature by the same terminology. In the literature, the
process of relativistic electrons scattering Langmuir waves
into transverse radiation is also called the electrostatic
bremsstrahlung \citep{Gailitis1964,Colgate1967, Melrose1971,
  Windsor1974, Akopyan1977, Schlickeiser2003}, which is
actually an induced scattering of transverse radiation off
of relativistic electrons mediated by Langmuir waves. The
``electrostatic bremsstrahlung'' of \cite{YZKS16} is the
emission of electrostatic eigenmodes by collisional process,
which is analogous to but distinct from the emission of
transverse electromagnetic radiation by collisional process.

As it will be demonstrated subsequently, the combined effects
of Langmuir wave-electron resonant interaction in the presence
of the new EB process leads to the self-consistent formation
of the core-halo EVDF, which is a process that may be operative
pervasively in the lower coronal environment. We thus suggest
that the present mechanism may be the most widely operative
process that is responsible for the formation of non-thermal
EVDFs, not only in the solar environment, but also in other
astrophysical environments. In the rest of this present Letter,
we detail the present finding.

\section{Theoretical formulation}
\label{sec:theor-form}
The essential idea behind the new process responsible for
taking a fraction of the electron population out of thermal
equilibrium and redistributing it to suprathermal velocity
and energy ranges is that the presence of the EB emission
term (as well as the collisional damping term) in the
wave-kinetic equation, combined with the particle kinetic
equation, leads to a new steady-state electron distribution
function, which corresponds to a Maxwellian core plus a
quasi-inverse power-law tail. Conceptually, such a state is
a new quasi-equilibrium that is distinct from thermodynamic
equilibrium. In such a state, enhanced electrostatic
fluctuations coexist with a population of charged particles
while maintaining a dynamical steady state. In order to
demonstrate this process, we numerically solve the system of
particle- and wave-kinetic equations of the generalized weak
turbulence theory \citep{YZKS16}, starting with an initially
Maxwellian electron velocity distribution and Langmuir wave
spectral intensity that reflects the presence of only the
customary spontaneous and induced emission, but not the EB
or the collisional damping. Of course, such an initial state
is out of force balance. The EB and collisional damping terms
subsequently drive the system to a new force-balanced steady
state for the wave intensity. The initial Langmuir fluctuation
spectrum is thus significantly modified as a result of the
additional terms in the wave-kinetic equation. The modified
Langmuir wave spectrum in turn distorts the initial Maxwellian
electron distribution, and transforms it into a new
quasi-steady-state velocity distribution that superficially
resembles the core-suprathermal velocity distribution function.
In what follows, we discuss the details of this numerical
demonstration.

We perform the self-consistent numerical analysis on the
EB emission in the Langmuir $(L)$ electrostatic eigenmode
frequency range by including the new mechanism in the
wave-kinetic equation. Instead of making use of the complete
set of nonlinear weak turbulence equations presented in
\cite{YZKS16}, we restrict our analysis to the quasi-linear
formalism, which includes single-particle spontaneous emission
and wave-particle induced emission. We also take into account
in the wave equations the effects of collisional damping. Such
a simplified approach allows the study of the time evolution
of the system, and puts in evidence the new mechanisms that
have been introduced in \cite{YZKS16}. Besides, in the absence
of free-energy sources, with which the present Letter is not
concerned, the nonlinear mode-coupling terms are not expected
to play any important dynamical roles. The equation describing
the dynamics of $L$ waves is therefore given by
\begin{eqnarray}
    \label{ILk}
  \frac{\partial I_{\bf k}^{\sigma L}}{\partial t}
  &=& \frac{\pi\omega_p^2}{k^2}\int d{\bf v}\,
    \delta\left(\sigma\omega_{\bf k}^L
    -{\bf k}\cdot{\bf v}\right)\\
  &&\times\left(\frac{n_0e^2}{\pi}\,F_e({\bf v})
    +\sigma\omega_{\bf k}^L\,I_{\bf k}^{\sigma L}
    \,{\bf k}\cdot\frac{\partial F_e({\bf v})}
    {\partial{\bf v}}\right)\nonumber\\
  &&+2\gamma_{\bf k}^{\sigma L}I_{\bf k}^{\sigma L}
    +P_{\bf k}^{\sigma L},\nonumber
\end{eqnarray}
where $I_{\bf k}^{\sigma L}$ is the wave intensity associated
with the Langmuir wave defined via $E_{{\bf k},\omega}^2
=\sum_{\sigma=\pm1}I_{\bf k}^{\sigma L}\delta(\omega
-\sigma\omega_{\bf k}^L)$, $E_{{\bf k},\omega}$ is
the spectral component of the wave electric field,
and the dispersion relation is given by
$\omega_{\bf k}^L=\omega_{pe}
\left(1+\frac{3}{2}k^2\lambda_D^2\right)$. Here,
$\omega_{pe}=\sqrt{4\pi n_0e^2/m_e}$ and
$\lambda_D=\sqrt{T_e/(4\pi n_0e^2)}$ stand for the
plasma frequency and Debye length, respectively, and
$n_0$, $e$, $m_e$, and $T_e$ are the ambient density,
unit electric charge, electron mass, and electron
temperature, respectively.

The first term on the right-hand side of Eq.~(\ref{ILk})
contains two contributions: the first term within the
large parenthesis, proportional to the EVDF, $F_e({\bf v})$,
represents the discrete-particle effect of spontaneous
emission; the second term, proportional to the derivative,
$\partial F_e({\bf v})/\partial{\bf v}$, represents the
induced emission. The second line of Eq.~(\ref{ILk}) on
the right-hand side includes the collisional wave damping
rate, $\gamma_{\bf k}^{\sigma L}$, obtained in the same context
as the EB \citep{YZKS16}, and numerically analyzed and
discussed in \cite{TZY16}. The collisional damping is
defined by
\begin{eqnarray}
\gamma_{\bf k}^{\sigma L} &=&
\omega_{\bf k}^L\frac{4n_ee^4\omega_{pe}^2}
{T_e^2}\int d{\bf k}'\frac{({\bf k}\cdot{\bf k'})^2
\lambda_D^4}{k^2{k'}^4|\epsilon({\bf k'},
\omega_{\bf k}^L)|^2}
\nonumber\\
&& \times\left(1+\frac{T_e}{T_i}
+({\bf k}-{\bf k'})^2\lambda_D^2\right)^{-2}
\nonumber\\
&& \times \int d{\bf v}\,{\bf k'}\cdot
\frac{\partial F_e({\bf v})}{\partial {\bf v}}
\delta(\omega_{\bf k}^L-{\bf k'}\cdot {\bf v}),
\label{gamma_coll}
\end{eqnarray}
where $T_i$ is the proton temperature, and
$\epsilon({\bf k},\omega)$ is the linear
dielectric-response function. In the literature, the
collisional damping rate of plasma waves are often computed
by heuristic means. That is, the collisional operator is
simply added to the exact Vlasov (or Klimontovich) equation
by hand, as it were, and the small-amplitude wave analysis
is carried out, leading to the so-called Spitzer formula for
the collisional damping rate \citep{Lifshitz1981}. A similar
heuristic and ad hoc recipe is also applied even for a
turbulent plasma \citep{Makhankov1968}. Such approaches are
at best heuristic and, strictly speaking, incorrect, as the
collisionality represents dissipation and irreversibility,
whereas the Vlasov or Klimontovich equation exactly preserves
the phase-space information, and thus is reversible. In the
non-equilibrium statistical mechanics it is well known that
the irreversibility enters the problem only as a result of
statistical averages and the subsequent loss of information.
The authors of \cite{YZKS16} carried out the rigorous analysis
of introducing the collisionality starting from the exact
Klimontovich equation and taking ensemble averages. The
collisional damping rate that emerged, namely
Eq.~(\ref{gamma_coll}), is the correct expression that
replaces the heuristic Spitzer formula, and it was found
in \cite{TZY16} that the heuristic Spitzer collisional
damping rate grossly overestimates the actual rate.

The term $P_{\bf k}^{\sigma L}$ in Eq.~(\ref{ILk}) describes
the electrostatic bremsstrahlung emission process, which is
new and is the subject of this Letter. In \cite{YZKS16} a
specific approximate form of $P_{\bf k}^{\sigma L}$ was derived.
In this Letter we have revisited the approximation procedure,
and we find that a more appropriate form is given by
\begin{eqnarray}
    \label{PkL,2}
  P_{\bf k}^{\sigma L}
  &&= \frac{3e^2}{4\pi^3}\frac{1}{(\omega_\bk^L)^2}
    \left(1-\frac{m_e}{m_i}\frac{T_e}{T_i}\right)^2\frac{v_e^4}{k^2}
    \int d{\bf k}'k'^2|\bk-\bk'|^2\nonumber\\
  &&\times\left(1+\frac{T_e}{T_i}
    +{k'}^2\lambda_D^2\right)^{-2}\left(1+\frac{T_e}{T_i}
    +({\bf k}-{\bf k}')^2\lambda_D^2\right)^{-2}\nonumber\\
  &&\times\int d{\bf v}\int d{\bf v}'
    \sum_{a,b}F_a({\bf v})F_b({\bf v}')\nonumber\\
  &&\times
    \delta[\sigma\omega_{\bf k}^L-{\bf k}\bm{\cdot}{\bf v}
    +{\bf k}'\bm{\cdot}({\bf v}-{\bf v}')],
\end{eqnarray}
where $m_i$ is the proton mass and $v_e=\sqrt{2T_e/m_e}$
stands for electron thermal speed. The detailed derivation of
the above-improved formula is reserved for another full-length
article, as it is too lengthy for the present Letter.

The dynamical equation for EVDF $F_e({\bf v})$
is given by the particle kinetic equation, which includes the 
Coulomb collision operator written in the form of velocity-space
Fokker-Planck equation,
\begin{eqnarray}
  \frac{\partial F_a({\bf v})}{\partial t}
  &=& \frac{\partial}{\partial v_i}
      \left(A_i({\bf v})\,F_a({\bf v})+D_{ij}({\bf v})\,
      \frac{\partial F_a({\bf v})}{\partial v_j}\right)
      \nonumber\\
  &&+\sum_b\theta_{ab}(F_a,F_b),\label{PKEq}\\
  A_i({\bf v})
  &=& \frac{e^2}{4\pi m_e}\int d{\bf k}
      \,\frac{k_i}{k^2}\sum_{\sigma=\pm1}\sigma
      \omega_{\bf k}^L\,\delta(\sigma\omega_{\bf k}^L
      -{\bf k}\cdot{\bf v}),
      \nonumber\\
  D_{ij}({\bf v})
  &=& \frac{\pi e^2}{m_e^2}\int d{\bf k}
      \,\frac{k_i\,k_j}{k^2}\sum_{\sigma=\pm1}
      \delta(\sigma\omega_{\bf k}^L
      -{\bf k}\cdot{\bf v})\,I_{\bf k}^{\sigma L},
      \nonumber
\end{eqnarray}
where the coefficient $A_i(\bv)$ represents the velocity-space
friction, and the coefficient $D_{ij}({\bf v})$ describes the
velocity diffusion. The distribution functions, $F_a({\bf v})$
and $F_b({\bf v})$, are both normalized to unity,
$\int d{\bf v} F_{a,b}({\bf v})=1$, where $a=e,i$ and $b=e,i$
represent the interacting particles.
The term $\theta_{ab}(F_{a},F_{b})$ depicts the effects of
Coulomb collisions between particles of species $a$ and $b$.

For the present analysis, we adopt a linearized
form of the Landau collision integral for $\theta_{ab}(F_a,F_b)$,
in which it is assumed that the evolving EVDF collides with
a Maxwellian background distribution. This assumption relies on
the fact that the growing tail population of the EVDF has a much
lower density than the core electrons, so that the effects of
collisions between the tail electrons with the background
EVDF are more significant than the effects of collisions among
electrons of the tail population. The lengthy linearization
procedure can be found in detail in \cite{TZYK16} and will
not be repeated here for the sake of space economy. In short,
the linearized collision operator is given by
\begin{eqnarray}
  \theta_{ab}
  &&(f_{a},f_{b}) = \Gamma_{ab}\left[
     \frac{\partial}{\partial{\bf v}_{a}}\cdot
     \left(2\frac{m_{a}}{m_{b}}\Psi(x_{ab})
     \frac{{\bf v}_{a}}{v^{3}_{a}}f_{a}\right)\right.
     \nonumber\\
  && +\frac{\partial}{\partial{\bf v}_{a}}\cdot
     \left\{\left[\left(\Phi(x_{ab})
     -\frac{1}{2x^{2}_{ab}}\Psi(x_{ab})\right)
     \frac{\partial^{2}v_a}{\partial{\bf v}_{a}
     \partial{\bf v}_{a}}\right]\cdot
     \frac{\partial f_{a}}{\partial{\bf v}_{a}}\right\}
     \nonumber \\
  && \left.+\frac{\partial}{\partial{\bf v}_{a}}\cdot
     \left[\left(\frac{1}{x^{2}_{ab}}\Psi(x_{ab})
     \frac{{\bf v}_{a}{\bf v}_{a}}{v^{3}_{a}}\right)\cdot
     \frac{\partial f_{a}}{\partial{\bf v}_{a}}\right]\right],
     \label{eq:coll-lin}
\end{eqnarray}
where $x_{ab}\equiv v_{a}/v_{tb}$, $v_{tb}$ is the thermal
velocity of the particles of species $b$,
$\Gamma=4\pi n e^{4} \ln\Lambda \, /m_{e}^{2}$, and
$\Psi(x)\equiv\Phi(x)-x\Phi'(x)$ is an auxiliary function
\citep{Gaffey1976}, in which
$\Phi(x_{ab}) \equiv \frac{2}{\sqrt{\pi}}
\int^{x}_{0}e^{-t^{2}}dt$ is the error function and
$\Phi'(x_{ab})=\frac{2}{\sqrt{\pi}}e^{-x^{2}}$ is its derivative.

\section{Numerical analysis}
\label{sec:numerical-analysis}
The set of integro-differential equations for waves and
particles, (\ref{ILk}) and (\ref{PKEq}) was numerically
solved in 2D wave-number space and 2D velocity space,
respectively. The purpose of the numerical analysis is to
demonstrate that the coupled system of equations leads to
an asymptotically steady-state EVDF that resembles the
core-halo distribution, regardless of how the solution is
initiated. As a concrete example, we assumed an initial
state of isotropic Maxwellian velocity distribution function
for both ions and electrons, given by
\begin{equation}
F_{a}({\bf v})=\frac{1}{\pi^{3/2}v_{a}^3}
\exp\left(-\frac{v^2}{v_{a}^2}\right),
\label{Fe_init}
\end{equation}
where $v_{a}=(2T_a/m_a)^{1/2}$, with $a=i,e$. The ion velocity
distribution is assumed to be constant along the time evolution,
which is a reasonable assumption as we are working in the
much-faster time-scale of electron interactions. The electron-ion
temperature ratio of $T_{e}/T_{i}=7.0$ is adopted, and the plasma
parameter of $\left(n_{0}\lambda_D^3\right)^{-1}=5\times10^{-3}$
is used. This choice represents a relatively high collisionality.
For the coronal-base source region, at the point where the plasma
becomes fully ionized, the electron density is of the order
$\unit[10^{9} - 10^{11}]{cm^{-3}}$ and the electron temperature may
reach $\sim \unit[10^4 - 10^6]{K}$ \citep{Aschwanden2005} [or
equivalently, $\sim \unit[10^0 - 10^2]{eV}$]. If we assume a central
value for the density and the temperature, $\unit[10^{10}]{cm^{-3}}$
and $\unit[10^{5}]{K}$, for instance, the corresponding plasma
parameter would be $\left(n_0\lambda_D^3\right)^{-1}\approx 10^{-5}$,
more than two orders of magnitude below the value, which we have used
for the numerical analysis. Such higher value was purposely utilized
in order to reduce the computational time necessary to obtain the
results. The final outcome of the time evolution, however, is not
affected by the inflated plasma parameter.

The initial Langmuir wave intensity was chosen by balancing
only the spontaneous- and induced-emission processes in the
equations for the wave amplitudes, namely,
\begin{equation}
I_{\bf k}^{\sigma L}(0)=\frac{T_e}{4\pi^2}
\left(1+3k^2\lambda_D^2\right).
\label{IkL_init}
\end{equation}
Because the velocity distribution and the Langmuir spectrum have
azimuthal symmetry, we plot the results of numerical solution by
using a 1D projection on the parallel direction of the velocity
and wave number.

It is important to note that the initial electron
distribution and Langmuir wave spectral intensity,
(\ref{Fe_init}) and (\ref{IkL_init}), {\it do not}
satisfy the steady-state condition $\partial/\partial t=0$
in the particle- and wave-kinetic Equations (\ref{PKEq})
and (\ref{ILk}), respectively. This is purposeful, since
our aim is to demonstrate the set of equations (\ref{PKEq})
and (\ref{ILk}) do not permit the electron distribution
and Langmuir wave spectral intensity, (\ref{Fe_init}) and
(\ref{IkL_init}), respectively, as the legitimate steady-state
solution, and so the equations will force the initial states
to make a transition to a new steady-state or, equivalently, a
new quasi-equilibrium state.

\begin{figure}
  \begin{center}
    \includegraphics[width=0.99\columnwidth]{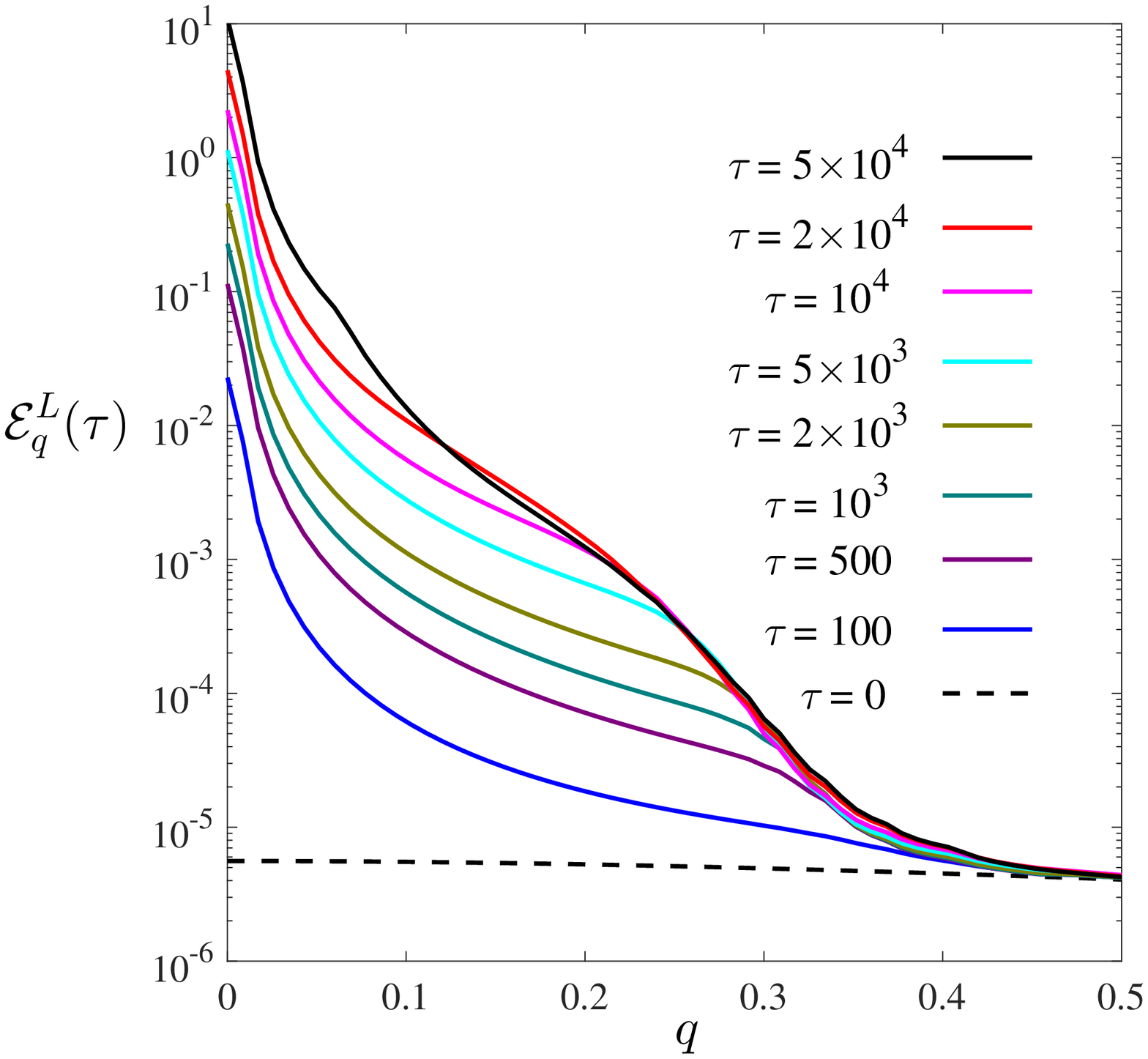}
    \caption{Time evolution of the Langmuir spectrum, taking into account
      the influence of the bremsstrahlung emission.}
    \label{fig1}
 \includegraphics[width=0.99\columnwidth]{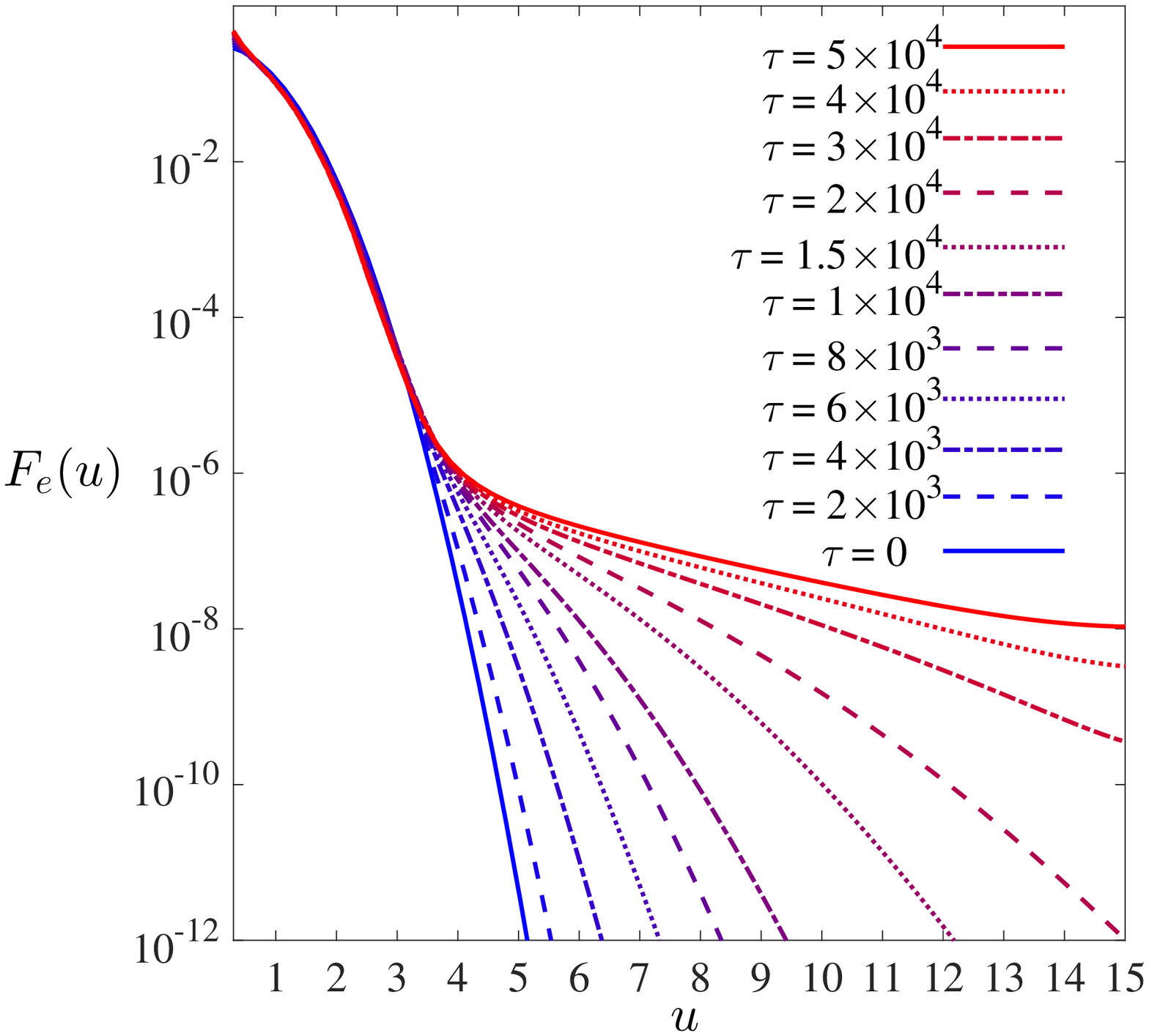}
    \caption{Time evolution of the electron velocity distribution function.}
    \label{fig2}
  \end{center}
\end{figure}

For the numerical analysis, we take into account the new
effects of collisional damping and EB, starting from the
above initial condition. With the addition of these new terms,
the initial wave spectrum is no longer in equilibrium with the
particle distribution, triggering an interesting evolution.
Let us define normalized Langmuir wave intensity
\begin{displaymath}
\mathscr{E}_{\bf q}^{\sigma L}=
\frac{(2\pi)^2 g}{m_ev_e^2}
I_{\bf k}^{\sigma\alpha},
\end{displaymath}
where $g=[2^{3/2}\,(4\pi)^2\,n_e\lambda_D^3]^{-1}$. We also define
the normalized temporal variable, $\tau=\omega_{pe}t$, and the
normalized wave number $kv_e/\omega_{pe}$. \autoref{fig1} shows
the time evolution of $\mathscr{E}_{\bf q}^{\sigma L}$. It is seen
that the bremsstrahlung radiation emitted in the frequency range
corresponding to $L$ waves alters the spectrum, creating a
modification that starts at $q\approx 0.4$ and ends in a peak
around $q=0$. The wave growth appears early in the time evolution
and evolves rapidly, as can be seen in \autoref{fig1}. After
$\tau=5000$, the shape of the curve starts to change and the wave
growth becomes slower. At $\tau=50,000$, the Langmuir spectrum
appears to be very close to an asymptotic state.

The early stages of the time evolution of the EVDF are quite
gradual, but the first signs of modification start to appear
around $\tau=2000$ and are almost imperceptible. In \autoref{fig2},
the earliest indication of change is shown at $\tau=4000$. At
this time an energized tail becomes apparent. The demarcation
between the core and tail occurs around $u=v/v_e\approx4.6$. The
velocity spectrum associated with the energetic tail population
continues to harden as time progresses, while the core defined for
$u\lesssim 4.6$ remains essentially unchanged.In short, we have
demonstrated that the initial Maxwellian electron distribution
(\ref{Fe_init}) has made a transition to a new quasi-equilibrium
state in which the electron distribution function bears a superficial
resemblance to the Maxwellian core plus a quasi-inverse power-law
tail population.

\section{Final remarks}
\label{sec:final-remarks}
The results obtained suggest that, in the presence of EB emission,
the wave-particle system attains a state of asymptotic equilibrium,
in which the EVDF possesses a feature of core-halo distribution
that is highly reminiscent of the solar wind EVDF. We thus conclude
that the present mechanism of the collective wave-particle
interaction process that takes place in a collisional environment,
such as the coronal source region, may be a highly efficient and
common process in many astrophysical environments. Before we close,
we note that we have also analyzed the particle kinetic equation in
which the collisional operator is not present on the right-hand side
of (\ref{PKEq}). The result (not shown) is not very different from
the present result, which indicates that the mechanism of generating
the suprathermal electrons mainly comes from the wave dynamics that
operate in a collisional environment.

We have also checked the overall energy  budget of the system.
Since the initial state, comprised of Maxwellian distribution
and Langmuir spectral intensity that does not reflect the
bremsstrahlung emission, is not in force balance, there is a
transfer of energy between the particles and waves early on,
but over a longer time period the system enters a state where
the net exchange between the particles and waves gradually
settles down to a minimal level. Note that in terms of the total
energy content, the tail portion of the EVDF contains a relatively
low proportion of the net energy, as the number density is several
orders of magnitude lower than the core distribution. Although it
is not so easy to verify by visual means, there is a slight cooling
associated with the core part of the EVDF. This shows that the
present process is not an acceleration mechanism, but rather it
involves the redistribution of particle population in velocity or
energy space in order to form a new quasi-equilibrium.

\vspace{-0.5cm}
\acknowledgments

S.F.T. acknowledges a PhD fellowship from CNPq (Brazil).
L.F.Z. acknowledges support from CNPq (Brazil), grant No. 304363/2014-6.
P.H.Y. acknowledges NSF grant AGS1550566 to the University
of Maryland, and the BK21 plus program from the National
Research Foundation (NRF), Korea, to Kyung Hee University.

\bibliographystyle{aasjournal}
\bibliography{papers}

\end{document}